\documentstyle[epsf]{mn}

\newif\ifAMStwofonts

\def\simlt{\lower.5ex\hbox{$\; \buildrel < \over \sim \;$}}
\def\simgt{\lower.5ex\hbox{$\; \buildrel > \over \sim \;$}}

\ifoldfss
  \ifCUPmtlplainloaded \else
    \NewTextAlphabet{textbfit} {cmbxti10} {}
    \NewTextAlphabet{textbfss} {cmssbx10} {}
    \NewMathAlphabet{mathbfit} {cmbxti10} {} 
    \NewMathAlphabet{mathbfss} {cmssbx10} {} 
  \fi
  \ifAMStwofonts
    \ifCUPmtlplainloaded \else
      \NewSymbolFont{upmath} {eurm10}
      \NewSymbolFont{AMSa} {msam10}
      \NewMathSymbol{\upi}     {0}{upmath}{19}
      \NewMathSymbol{\umu}     {0}{upmath}{16}
      \NewMathSymbol{\upartial}{0}{upmath}{40}
      \NewMathSymbol{\leqslant}{3}{AMSa}{36}
      \NewMathSymbol{\geqslant}{3}{AMSa}{3E}

    \fi
  \fi
\fi 

\ifnfssone
  \newmathalphabet{\mathit}
  \addtoversion{normal}{\mathit}{cmr}{m}{it}
  \addtoversion{bold}{\mathit}{cmr}{bx}{it}
  \newmathalphabet{\mathbfit} 
  \addtoversion{normal}{\mathbfit}{cmr}{bx}{it}
  \addtoversion{bold}{\mathbfit}{cmr}{bx}{it}
  \newmathalphabet{\mathbfss} 
  \addtoversion{normal}{\mathbfss}{cmss}{bx}{n}
  \addtoversion{bold}{\mathbfss}{cmss}{bx}{n}
  \ifAMStwofonts
    \ifCUPmtlplainloaded \else
      \UseAMStwoboldmath
      \makeatletter
      \new@mathgroup\upmath@group
      \define@mathgroup\mv@normal\upmath@group{eur}{m}{n}
      \define@mathgroup\mv@bold\upmath@group{eur}{b}{n}
      \edef\UPM{\hexnumber\upmath@group}
      \new@mathgroup\amsa@group
      \define@mathgroup\mv@normal\amsa@group{msa}{m}{n}
      \define@mathgroup\mv@bold\amsa@group{msa}{m}{n}
      \edef\AMSa{\hexnumber\amsa@group}
      \makeatother
      \mathchardef\upi="0\UPM19
      \mathchardef\umu="0\UPM16
      \mathchardef\upartial="0\UPM40
      \mathchardef\leqslant="3\AMSa36
      \mathchardef\geqslant="3\AMSa3E
    \fi
  \fi
\fi 

\ifnfsstwo
  \DeclareMathAlphabet{\mathbfit}{OT1}{cmr}{bx}{it}
  \SetMathAlphabet\mathbfit{bold}{OT1}{cmr}{bx}{it}
  \DeclareMathAlphabet{\mathbfss}{OT1}{cmss}{bx}{n}
  \SetMathAlphabet\mathbfss{bold}{OT1}{cmss}{bx}{n}
  \ifAMStwofonts
    \ifCUPmtlplainloaded \else
      \DeclareSymbolFont{UPM}{U}{eur}{m}{n}
      \SetSymbolFont{UPM}{bold}{U}{eur}{b}{n}
      \DeclareSymbolFont{AMSa}{U}{msa}{m}{n}
      \DeclareMathSymbol{\upi}{0}{UPM}{"19}
      \DeclareMathSymbol{\umu}{0}{UPM}{"16}
      \DeclareMathSymbol{\upartial}{0}{UPM}{"40}
      \DeclareMathSymbol{\leqslant}{3}{AMSa}{"36}
      \DeclareMathSymbol{\geqslant}{3}{AMSa}{"3E}
    \fi
  \fi
\fi 

\ifCUPmtlplainloaded \else
  \ifAMStwofonts \else 
    \def\upi{\pi}
    \def\umu{\mu}
    \def\upartial{\partial}
  \fi
\fi

\title[Constraining the window]
	{Constraining the window on sterile neutrinos as warm dark matter}
\author[Hansen, Lesgourgues, Pastor \& Silk]
{Steen H. Hansen (1), Julien Lesgourgues (2), Sergio Pastor (3) and Joseph Silk(1)\\
(1) Department of Physics, Nuclear \& Astrophysics Laboratory,
University of Oxford, Keble Road, Oxford OX1 3RH, U.K.\\
(2) Laboratoire de Physique Th\'eorique LAPTH, B.P. 110, F-74941,
Annecy-le-Vieux Cedex, France\\
(3) Max-Planck-Institut f\"{u}r Physik 
(Werner-Heisenberg-Institut),
F\"{o}hringer Ring 6, D-80805, Munich, Germany}
\date{Draft version \today}
\pagerange{\pageref{firstpage}--\pageref{lastpage}}
\pubyear{2001}

\begin{document}

\maketitle

\label{firstpage}

\begin{abstract}
Sterile neutrinos may be one of the best Warm Dark Matter candidates
we have today. Both lower and upper bounds on the mass of the sterile
neutrino come from astronomical observations. We show that the proper
inclusion of the neutrino momentum distribution reduces the allowed
region to be $2.6$ keV $< m< 5$ keV for the simplest models.  A search
for a spectral line with $E=m/2$ is thus more interesting than ever
before.
\end{abstract}




\section{Introduction}
\label{secint}
Astrophysics provides an increasing amount of independent indications
that the dark matter of the universe is warm, so that the small-scale
fluctuations are damped out by free streaming.  This is most easily
achieved by giving a keV mass to the DM particle, in which case the
preferred candidate is the sterile neutrino.  Support for warm dark
matter (WDM) comes from simulations of the number of satellite
galaxies (Colin, Avila-Reese \& Valenzuela 2000) and of disk galaxy
formation without the need for stellar feedback (Sommer-Larsen \&
Dolgov 2000), which both find that a DM particle mass of about 1 keV
is optimal: a significantly larger mass has little impact on galaxy
formation, and a significantly smaller mass would lead to the well
known difficulties faced by hot dark matter. A quantitative lower
limit on the candidate WDM particle mass is inferred from the
existence of a massive black hole at large redshift (Barkana, Haiman
\& Ostriker 2001) and the requirement of sufficiently early galaxy
formation to account for reionization of the universe and the observed
Ly-$\alpha$ forest properties (Narayanan et al. 2000), constraining
the DM mass to be larger than 0.75 keV. A recent discussion of x-ray
emission from decays of sterile neutrinos (Abazajian, Fuller \& Tucker
2001) has imposed an upper limit of about 5 keV on the neutrino mass.
Here we discuss a reinterpretation of these bounds on neutrino mass,
and demonstrate that the proper inclusion of the neutrino momentum,
arising from the specific production temperature, reduces the allowed
sterile neutrino WDM mass to be in the range $2.6$ keV $< m< 5$ keV.

\section{WDM particle decoupling}

All of these studies (Colin et al. 2000; Sommer-Larsen \& Dolgov 2000;
Barkana et al. 2001; Narayanan et al. 2000) are based on the
mass-dependent cut-off on small scales, produced by free-streaming. In
the previously cited studies, a ``conventional'' WDM model was
considered for the underlying particle physics (see Bode, Ostriker \&
Turok (2001) and Sommer-Larsen \& Dolgov (2000) for recent overviews
of such particle models). In such conventional WDM models, the
particles decouple in the early Universe at higher temperatures than
do massless neutrinos.  Therefore they do not share the entropy
release from the successive particle annihilations. Since they were
relativistic at decoupling, their distribution function in momentum
space is subsequently that of a massless fermion, but with a
temperature, $T_W$, which is given today by
\begin{equation}
T_{W_0} = T_{\nu_0} \, \left( \Omega_W h^2 \, 
\frac{94 \, \mbox{eV}}{m_W}\right)^{1/3} \, ,
\end{equation}
where $T_{\nu_0}\approx1.946$ K, $H_0 = 100 \, h \mbox{km} \,
\mbox{s}^{-1} \mbox{Mpc}^{-1}$, $\Omega_W$ is the present energy
density of WDM in units of the critical density, and $m_W$ is the WDM
mass. The needed entropy release in these conventional models is much
bigger than allowed in the standard model, and such WDM candidates
should therefore have decoupled before a larger gauge group breaks
down.

Now a very natural candidate for the WDM particle is a massive sterile
neutrino mixed with an ordinary neutrino (Dodelson \& Widrow 1994;
Colombi, Dodelson \& Widrow 1996; Shi \& Fuller 1999; Dolgov \& Hansen
2001; Abazajian, Fuller \& Patel 2001). Since the mixing angle is
temperature dependent (N\"otzold \& Raffelt 1988) (and for small
vacuum mixing angle, $\sin^22\theta \sim 10^{-7}$), only a small
amount of these heavy neutrinos, relative to ordinary active
neutrinos, can be produced at high temperatures. The distribution
function of sterile neutrinos is, to a fair approximation,
characterized by the temperature of massless neutrinos, but smaller by
a factor ${\cal X}$. For a specific choice of $m_W$ and $\Omega_W$
today the value of ${\cal X}$ can be found from
\begin{equation}
{\cal X} = \Omega_W h^2 \, \left( \frac{94 \, \mbox{eV}}{m_W} \right) 
\sim 10^{-2} \, ,
\end{equation}
therefore the two models produce the same contribution to $\rho_{tot}$
of WDM particles today if
\begin{equation}
{\cal X} =   \left( \frac{T_{W_0}}{T_{\nu_0}} \right)^3 \, .
\end{equation}
However, WDM particles with $m_W$ have a {\it different} distribution
function in these two models, and their free-streaming effect is not
equivalent. The effect on large scale structure was first discussed in
detail by Colombi et al. (1996).  We will use conventional WDM (cWDM)
when referring to the first case, and sterile neutrino WDM (sWDM) for
the second one.

\section{Lower bounds}
The two neutrino models are easily included in a Boltzman code, in
order to compute the present matter power spectrum $P(k)$. Using the
code {\it cmbfast} (Seljak \& Zaldarriaga 1996), we have found
analytical fits for the transfer functions, $T(k)$, relating the power
spectrum in the WDM to the CDM scenario
\begin{equation}
T^2(k) = \frac{P^W (k)}{P^{CDM} (k)} \, \, \, \, \, 
\mbox{for} \, \, \, \, \,  \Omega_W = \Omega_{CDM} \, ,
\end{equation}
where $P^W$ is the power spectrum for cWDM, and a similar expression
with $P^{\nu}(k)$ for the sWDM model.  These transfer functions, which
essentially reflect the free streaming cut-off, have the form
\begin{equation}
T(k) = \left[ 1 + \left( \alpha k\right)^{2 \nu} \right] ^{-5/\nu} \, ,
\end{equation}
where $k$ is the wavenumber in units $h \mbox{Mpc}^{-1}$, $\nu=1.12$,
and $\alpha$ depends on the cosmological parameters as
\begin{equation}
\alpha = A \left(\frac{\Omega_W}{0.3} \right) ^b 
\left(\frac{h}{0.65} \right) ^c   
\left(\frac{m_W}{500 \, \mbox{eV}} \right) ^d \, .
\label{alph}
\end{equation}
Numerically we find for cWDM, $A=1.07$, $b=0.11$, $c=1.20$ and
$d=-1.11$ in good agreement with Bode et al. (2001). For the sWDM one
can derive similar numbers by noting that the mass in the cWDM case
differs by $(T_{W_0}/T_{\nu_0} )$. This means that if we have a
dependence $\Omega_W^b h^c m_W^d$ for the sWDM case in
eq.~(\ref{alph}), and a dependence $\Omega_W^{b' }h^{c'} m_W^{d'}$ for
the cWDM case, then one finds
\begin{equation}
\Omega_W ^b h^c m_W^d = \Omega_W^{b'-d'/3} h^{c'-2d'/3} m_W^{d'+d'/3} \, ,
\end{equation}
which is solved by $b'=b+d/4$, $d'=3d/4$ and $c'=c+d/2$, in good
agreement with what we found numerically, by explicitly changing the
massive neutrino phase-space distribution function as done by
Lesgourgues \& Pastor (1999).  To be very explicit, this means that
for a given cut-off scale of the power spectrum one can find the
corresponding mass of the elementary particle, and the mass will
differ in the two cases. E.g. if for cWDM one finds $m_W=0.75$ keV,
then this corresponds in the sWDM case to $m_W = (\Omega_W h^2 \, 94
\mbox{eV}/0.75 \mbox{keV})^{-1/3} \cdot 0.75 \mbox{keV} \approx 2.6$
keV, when using $h=0.7$ and $\Omega_W=0.4$.  In other words, if one
believes that sterile neutrinos indeed constitute the dark matter, and
they are produced as described in (Dodelson \& Widrow 1994; Colombi,
Dodelson \& Widrow 1996; Dolgov \& Hansen 2001; Abazajian, Fuller \&
Patel 2001), then the bounds obtained by Barkana et al. (2001) and
Narayanan et al.  (2000), $m_W > 0.75 \, \mbox{keV}$, should really be
multiplied by a factor $3.4$, and the {\it lower} bound on sterile
neutrinos as dark matter is thus about $2.6$ keV.

This lower bound may be subject to minor corrections. First, the
temperature of the sterile neutrinos is really slightly lower than the
active neutrino temperature, since the sterile neutrinos are being
produced while the muons are still present in the Universe, $T\approx
130$ MeV (Langacker 1989; Kainulainen 1990; Barbieri \& Dolgov 1990,
1991).  Similarly, for the neutrino states being produced above the
QCD phase transition, one must also take into account the quark
degrees of freedom (Abazajian et al. 2001a).  Another effect arises
from the fact that the produced sterile neutrino spectrum is not
exactly thermal, but slightly warmer in the sense that the higher
momentum part is more populated than the lower momentum part (Dolgov
\& Hansen 2001; Abazajian et al. 2001b).  Furthermore, the factor of
3-4 found above depends on the specific values of $\Omega_W$ and $h$,
and can therefore change slightly.  It is also worth noting, that if
the sterile neutrinos are produced resonantly (Shi \& Fuller 1999),
through a pre-existing lepton asymmetry, then the upper limit on 5 keV
may weaken substantially (Abazajian et al. 2001b).

\section{Upper bounds and discussion}
Sterile neutrinos in the keV mass range have a decay time that is of
cosmological interest.  Recently a very interesting paper appeared
(Abazajian et al. 2001b), where the signature from decaying sterile
neutrinos in galaxies and clusters of galaxies was studied in
detail~\footnote{One could imagine a slight change of strategy in the
analysis of (Abazajian et al. 2001b), namely to consider regions with
large concentration of dark matter, but with little baryonic
matter. Such ``dark blobs'' may have been observed by inverting the
matter distribution in clusters of galaxies from weak lensing, and
comparing with the baryonic matter inferred from optical observations
of the cluster (Clowe et al. 2000), but the significance of such blobs
is far from being established.}.  A bound $m<5$ keV was derived, using
the relation between mass and mixing angle obtained in (Abazajian et
al. 2001a).  We thus conclude that a sterile neutrino as WDM must lie
in the mass range $2.6$ keV $< m< 5$ keV, and it is therefore more
interesting than ever before to search for a spectral line with energy
$E=m/2$ from the decay $\nu_s \rightarrow \nu_\alpha + \gamma$.

If future searches for a spectral line from the sterile neutrino decay
should give a negative result, then one must find new WDM
candidates. An interesting possibility is an active neutrino, which
may never reach thermal equilibrium if the reheat temperature at the
end of inflation is low enough (Giudice et al. 2001).  Such a scenario
demands a reheat temperature of a few MeV, and the resulting neutrino
distribution function is also warmer than a conventional WDM
candidate.  Such a solution does, however, not appear too natural in
view of the recent neutrino data (Ahmad et al. 2001; Toshito et
al. 2001), indicating that all the active neutrino masses are sub eV.
Several WDM candidates have different bounds from what is presented
here, e.g.  gravitinos (Pagels \& Primack 1982; Borgani, Masiero \&
Yamaguchi 1996; Kawasaki, Sugiyama \& Yanagida 1997) produced in the
very early Universe; or sterile neutrinos {\em if} there should exist
an initial lepton asymmetry (Shi \& Fuller 1999).  Even more
traditional DM candidates (like a neutralino) can disguise themselves
as WDM, namely if they have scattering cross section with photons or
neutrinos.  Specifically, for a 100 GeV DM particle with scattering
cross section about $\sigma_{DM-\gamma} \approx 10^{-30} \mbox{cm}^2$
one obtains a reduction of the matter power spectrum corresponding to
a conventional WDM candidate with mass about 1 keV (B\oe{}hm et
al. 2001).

In conclusion, we have shown that the proper inclusion of the neutrino
momentum distribution changes the lower bound allowed for the simplest
models of sterile neutrinos as WDM, and the resulting allowed region
thus becomes $2.6$ keV $< m< 5$ keV.

\section*{Acknowledgement}
It is a pleasure for SHH and SP to thank Kev Abazajian, George Fuller,
Kimmo Kainulainen and Dima Semikoz for discussions and comments.  SHH
and SP are supported by Marie Curie Fellowships of the European
Commission under contracts HPMFCT-2000-00607 and HPMFCT-2000-00445.
They acknowledge a visit to LAPTH, supported by CNRS, during which
part of this work was carried out.

\label{lastpage}

\end{document}